\documentclass[aps,pre,twocolumn,groupedaddress,showpacs]{revtex4}

\usepackage{amsmath}
\usepackage{amssymb}
\usepackage{graphicx}% Include figure files

\begin{document}

%\preprint{}

\title{Dynamics of a Brownian circle swimmer}
\author{Sven van Teeffelen}
\email[]{teeffelen@thphy.uni-duesseldorf.de}
\author{Hartmut L\"owen}
 \affiliation{Institut f\"ur Theoretische Physik II: Weiche Materie,
Heinrich-Heine-Universit\"at D\"usseldorf, 
D-40225 D\"usseldorf, Germany}
\date{\today}

\begin{abstract}
  Self-propelled particles move along circles rather than along a
  straight line when their driving force does not coincide with their
  propagation direction. Examples include confined bacteria and
  spermatozoa, catalytically driven nanorods, active, anisotropic
  colloidal particles and vibrated granulates.  Using a
  non-Hamiltonian rate theory and computer simulations, we study the
  motion of a Brownian ``circle swimmer'' in a confining channel.  A
  sliding mode close to the wall leads to a huge acceleration as
  compared to the bulk motion, which can further be enhanced by an
  optimal effective torque-to-force ratio.
\end{abstract}
\maketitle
Active particles, which are self-propelled by their own motor, exhibit a
wealth of novel and fascinating nonequilibrium effects such as giant
density fluctuations~\cite{Ramaswamy:02}, swarming~\cite{Vicsek:95},
and swirling~\cite{Kudrolli:07}.  Examples are found in quite
different areas of physics and include micro-organisms propelled by
flagella in a
fluid~\cite{Riedel:05,Woolley:03,DiLuzio:05,Hill:07,Berg:90}, man-made
colloidal swimmers~\cite{Dreyfus:05}, catalytically driven nanorods or
Janus particles~\cite{Dhar:06,Walther:08}, vibrated granulates of
polar rods~\cite{Kudrolli:07,Narayan:07}, and
pedestrians~\cite{Helbing:01}.  Typically it is assumed that the
swimmers move along their symmetry axis such that the force and the
particle orientation are in line.  This leads to a motion along a
straight line just perturbed by random (e.g., Brownian) fluctuations.

Here we study the case in which the internal force propelling a colloidal
particle does not coincide with the particle orientation. In the
absence of Brownian fluctuations, this will lead to an overdamped
motion along a closed circle, therefore we refer to this particle as a
``circle swimmer.''  Even a slight misalignment of the drive direction
will result in circle swimming, which is thus the generic case of
self-propulsion.  Circle swimmers with a pronounced curved trajectory
are realized in nature and can be artificially prepared: In fact, it
has been shown that certain
bacteria~\cite{Berg:90,DiLuzio:05,Lauga:06,Hill:07} and
spermatozoa~\cite{Riedel:05,Woolley:03}, when confined to two
dimensions, swim in circles. Moreover, catalytically driven
nanorods~\cite{Dhar:06,Walther:08} and colloidal
particles~\cite{Dreyfus:05} can be prepared with a tilted motor, and a
vibrated polar rod~\cite{Kudrolli:07} on a planar substrate with an
additional left-right asymmetry will move along circles. Last but not
least, the trajectories of completely blinded and ear-plugged
pedestrians have a significant circular form~\cite{Obata:05}.  Despite
their practical importance, the Brownian dynamics of a circle swimmer
has not yet been addressed by theory and simulation either in the bulk
or under confinement~\cite{footnote_swimmer_spiral}.

In this paper, we propose a simple model for Brownian motion of a
circle swimmer in two spatial dimensions arising from the combined
actions of an internal self-propelling force and a torque.  We solve
the Langevin equation of a two-dimensional circle swimmer analytically
in the bulk providing a suitable reference model.  The averaged
position falls on a {\it spira mirabilis}, and a crossover from an
oscillatory ballistic to a diffusive behavior is found in the
mean-squared displacement. We then identify the modes of propagation
of a circle swimmer in confining channels with repulsive walls using
computer simulations and a non-Hamiltonian rate theory.  In symmetric
channels, the long-time self-diffusion coefficient $D_L$ is
significantly enhanced mediated by an efficient {\it sliding mode} of
a tilted rod close to a wall.  Furthermore, $D_L$ is nonmonotonic in
the torque.  Finally, in asymmetric channels which are lacking a
left-right symmetry (e.g., due to gravity~\cite{Koppl:06}), the
sliding mode of the circle swimmer yields a {\it ballistic} motion
along the wall.

\begin{figure}
  \includegraphics[width=8cm]{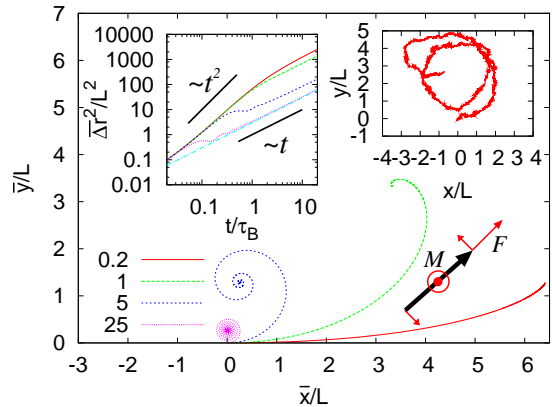}
  \caption{(Color online) Trajectories of the mean position $\overline{{\bf r}}$ of
    the self-propelling rod for fixed $\beta FL=10$, $\beta
    M=0.2,1,5,25$ (${\bf r}_0={\bf 0}, \phi_0=0$). Left inset: the
    mean-square displacement $\overline{\Delta {\bf r}^2}$ for the
    same force and torques, but also for $\beta FL=0$, $\beta M=0$
    (lowermost curve). Right inset: a typical trajectory of the rod
    for $\beta FL=25$, $\beta M=10$, for times $0<t<\tau_B$. Lower
    right inset: Sketch of the self-propelled circle swimmer.}
  \label{fig:free_rbar_trajectory_msd}
\end{figure}

Neglecting hydrodynamic interactions, the overdamped motion of the
Brownian circle swimmer in two dimensions is governed by the Langevin equations
for the rod center-of-mass position $\dot{\bf r}= \beta{\bf D}\cdot \left[
F{\bf \hat u}-{\bf \nabla} V({\bf r},\phi)+ {\bf f}\right]$ and for the rod
orientation $\dot \phi= \beta D_r \left[M -
\partial_\phi V({\bf r},\phi)+ \tau\right]$, respectively, where dots denote
time derivatives and $\beta^{-1}=k_BT$ is the thermal energy.  The
rod's short time diffusion tensor ${\bf D}=D_\parallel({\bf \hat
  u}\otimes{\bf \hat u})+D_\perp({\bf I}-{\bf \hat u}\otimes{\bf \hat
  u})$ is given in terms of the short time longitudinal
($D_\parallel$) and transverse ($D_\perp$) translational diffusion
constants, with ${\bf \hat u}=(\cos\phi,\sin\phi)$, ${\bf I}$ the unit
tensor and $\otimes$ a dyadic product.  $D_r$ is the short time
rotational diffusion constant.  $F{\bf \hat u}$ is a constant {\it
  effective} internal force that represents the propulsion mechanism
responsible for the deterministic motion in the rod orientation, and
$M$ is a constant {\it effective} internal or external torque yielding
the deterministic circular motion (see the sketch in
Fig.~\ref{fig:free_rbar_trajectory_msd}). $V({\bf r},\phi)$ is an
external confining potential. ${\bf f}$ and $\tau$ are the zero mean
Gaussian white noise random force and random torque originating from
the solvent, respectively. Their variances are given by
$\overline{f_\parallel(t)f_\parallel(t^\prime)} =
2\delta(t-t^\prime)/(\beta^2 D_\parallel)$,
$\overline{f_\perp(t)f_\perp(t^\prime)} = 2\delta(t-t^\prime)/(\beta^2
D_\perp)$, and $\overline{\tau(t)\tau(t^\prime)} =
2\delta(t-t^\prime)/(\beta^2 D_r)$, where $f_\parallel$, $f_\perp$ are
the components of ${\bf f}$ parallel and perpendicular to $\bf \hat
u$, respectively. The bars over the quantities denote a noise average.
We remark that for an active self-propelled particle, $F$ and $M$ are
{\it effective} net forces that could be determined in the bulk from
the forward and angular velocities $F=|\dot {\vec r}|/(\beta
D_\parallel)$ and $M=|\dot \phi|/(\beta D_r)$, respectively, but are
not necessarily directly connected to the internal propulsion
mechanism~\cite{Raz:07}.

At first we consider the free circle swimmer, i.e., we set $ V({\bf
  r},\phi)=0$. In the limit of zero temperature, the rod center of mass
would describe a perfect circle of radius $R=(D_\parallel F)/(D_r M)$,
with the circular frequency $\omega\equiv\beta D_rM$. For finite
temperature all moments of ${\bf r}$ and $\phi$ can be calculated
exactly.  The first and second moments of $\phi(t)$ are simply given
by $\overline{\phi}= \phi_0+\omega t$ and $\overline{\Delta\phi^2} =
\overline{\left[\phi(t)-\phi_0\right]^2} = (\omega t)^2+2D_r t$, where
$\phi_0=\phi(t=0)$, and where we let $\phi$ run ad infinitum.  The
first two moments of $\Delta{\bf r}\equiv{\bf r}(t)-{\bf r}(0)$ are
given by
\begin{align}\begin{split}\label{eq:moments_r}
\overline{\Delta{\bf r}}
=&\lambda \Big[D_r{\bf \hat  u}_0 + \omega {\bf \hat  u}_0^\perp -
e^{-D_rt}\left(D_r{\bf \overline{\hat{ u}}}+\omega{\bf \overline{\hat{ u}}}^\perp\right)\Big]\\
\overline{\Delta{\bf r}^2}
%\equiv\left<\left[{\bf r}(t)-{\bf r}(0)\right]^2\right>
=&2\lambda^2 \Big\{\omega^2-D_r^2+D_r(D_r^2+\omega^2)t\\
  &+e^{-D_r t}\left[(D_r^2-\omega^2)\cos(\omega t)-2 D_r \omega\sin(\omega t)\right]\Big\}\\
&+2(D_\parallel+D_\perp)t\,,
\end{split}\end{align}
with $\lambda=\beta D_\parallel F/(D_r^2+\omega^2)$, ${\bf \hat u}_0 =
(\cos\phi_0, \sin\phi_0)$, ${\bf \hat u}_0^\perp = (-\sin\phi_0,
\cos\phi_0)$, ${\bf \overline{\hat{ u}}} = (\cos\overline{\phi},
\sin\overline{\phi})$, and ${\bf \overline{\hat{ u}}}^\perp =
(-\sin\overline{\phi}, \cos\overline{\phi})$, i.e., $\overline{\Delta
  {\bf r}}$ describes a {\it spira mirabilis}.

We consider a very thin rod of length $L$, where $D_r/D_\parallel=3/(2
L^2)$, $D_\perp=D_\parallel/2$. We will denote all times in units of
$\tau_B=L^2/D_\parallel$, lengths in units of $L$, and energies in
units of $\beta^{-1}$. Different regimes are distinguished in terms of
the dimensionless quantities $D_r/\omega$ and $\beta FL$.  The latter
determines whether the rod's erratic motion is dominated by the kicks
of the solvent particles or by the self-propulsion. The former is the
ratio of the ballistic over the random turning rate.  In
Fig.~\ref{fig:free_rbar_trajectory_msd}, we show $\overline{\Delta {\bf
    r}}$ for different internal torques $M$ and a typical trajectory
of the rod position during two complete turns.
In the second inset of Fig.~\ref{fig:free_rbar_trajectory_msd} we
display $\overline{\Delta {\bf r}^2}$, which shows deterministic
behavior for $t\lesssim 1/{D_r}$ while for large times the swimmer
moves in a random fashion according to $\overline{\Delta {\bf
    r}^2}\propto t$.

\begin{figure}
  \includegraphics[width=7cm]{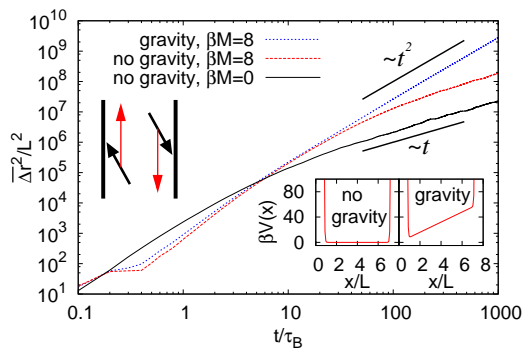}
  \caption{(Color online) Mean-square displacement $\overline{\Delta{\bf r}^2}$ in
    confinement ($\beta FL=60$, $L_x=8L$) without gravity and with
    zero torque (black), without gravity and with finite torque (red),
    and with torque and gravity (blue). The left inset displays the
    rod sliding along the walls. The right inset shows the confining
    potential without (left) and with (right) gravity.}
  \label{fig:msd_confinement_gravity}
\end{figure}

Next, we introduce a confining, integrated segment-wall power-law
potential in the $x$ direction, $V(x,\phi)=\int_0^L{\mathrm
  d}l\,v\left[x^\prime(l)\right]+kx$ with $v(x^\prime)\equiv(\beta L)^{-1}
\{[L/x^\prime]^n + [L/(L_x-x^\prime)]^n\}$, where $L_x$ is the channel
width, $n=24$ is a large exponent, and $x^\prime(l)$ is the $x$ position of
the rod segment at contour length $l$ (see the right inset of
Fig.~\ref{fig:msd_confinement_gravity}).  In case the solvent is
confined as well, hydrodynamic interactions between the particle and
the wall lead in principle to an $x$-dependent diffusion
tensor~\cite{Grier:00}, which is ignored in our model. An additional
gravitational force in the $x$ direction~\cite{Koppl:06} of strength
$k$ will be applied later, but we focus first on the symmetric case
$k=0$.  At zero temperature, for a not too large ratio $M/LF$ and
under appropriate initial conditions (${\bf r}_0,\phi_0$), the tilted
swimmer performs a steady-state {\it sliding motion} along either of
the two walls with a constant $x$-position close to the wall and with
a constant angle $\phi$ determined by the steady-state conditions
$\dot{x}=0$, and $\dot{\phi}=0$, respectively.  Without loss of
generality, we consider the case $M>0$, i.e., the rod rotates
counterclockwise, such that it slides upwards along the left wall
(see the sketch in Fig.~\ref{fig:msd_confinement_gravity}).  In the limit
of hard walls ($n\rightarrow\infty$), the two solutions to the set of
steady-state equations can be given explicitly as $x_{s/u}=L(1-1/2
\cos\phi_{s/u})$, (i.e., the front rod tip sits on the wall), and
$\cos^2\phi_{s/u}=[1-2(M/LF)^2\mp\sqrt{1-8(M/LF)^2}]/[2+2(M/LF)^2]$,
$\cos\phi_{s/u}<0$, where the minus sign corresponds to the stable
($\phi_s$) and the plus sign to the unstable ($\phi_u$) solution.
Clearly, for $2\sqrt{2}M/LF>1$ there is no solution to the
steady-state conditions, but the rod keeps on rotating. For large
exponents $n$, the asymptotic steady-state velocity in the
$y$ direction is given by $v_y \simeq D_\parallel
F\sin\phi_s/(1+\cos^2\phi_s)$.

The sliding mode is also present at finite temperature.  However, by
thermal fluctuations the rod eventually leaves the wall and reaches
the opposite wall under an appropriate angle for the respective
sliding mode in the opposite $y$ direction, which we refer to as
``flipping.''  Consequently, the circle swimmer moves diffusively
according to $\overline{\Delta {\bf r}^2}\simeq 2 D_L t$, with $D_L$
the long-time translational diffusion coefficient. This picture is
clearly confirmed by Brownian dynamics computer simulations, averaged
over $1000$ independent simulation runs, as shown in
Fig.~\ref{fig:msd_confinement_gravity}.

\begin{figure}
  \includegraphics[width=7cm]{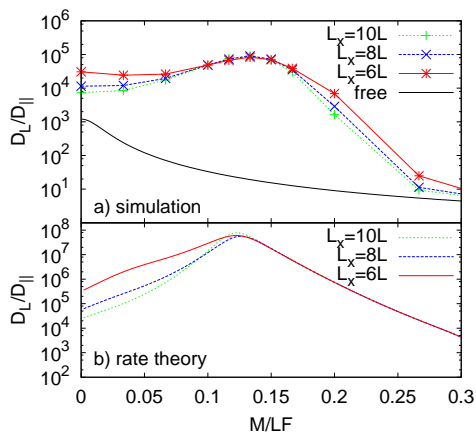}
  \caption{(Color online) Long-time diffusion coefficient $D_L$ as a function of the
    torque $M$ for the swimmer in the bulk and in confinement for
    $\beta FL=60$ and $L_x=6,8,10L$.  (a) Computer simulation, (b) rate
    theory.}
  \label{fig:longtimediffusion_confinement_bulk}
\end{figure}

For large $\beta FL$, large $\beta M$, and a channel width of the
order of the circle radius ($L_x\lesssim R$), the average time the
swimmer spends in its stable mode on either of the walls is large as
compared to the duration of a flip.  Thus, the swimmer effectively
performs a one-dimensional random walk with a typical step length
$a\simeq v_y/\gamma$, where $\gamma$ is the flipping rate. This random
walk leads to a long-time diffusion coefficient of $D_L\simeq
v_y^2/\gamma$, which we display as a function of internal torque $M$
for different wall-wall separations $L_x$ in
Fig.~\ref{fig:longtimediffusion_confinement_bulk}.

It is clearly seen from the simulations
[Fig.~\ref{fig:longtimediffusion_confinement_bulk}(a)] that the
diffusion in the channel is strongly enhanced as compared to the
diffusivity of the free swimmer. In particular, this strong
enhancement is already observed for $M=0$, as the narrow walls
constantly align the rod in the $y$ direction. However, the diffusion
eventually slows with increasing wall-wall separation $L_x$.  For
intermediate $M/LF\approx 0.15$, diffusion is enhanced even
further---in the simulations
[Fig.~\ref{fig:longtimediffusion_confinement_bulk}(a)] by an order of
magnitude---displaying a much smaller dependence on $L_x$. This
non-monotonic behavior of $D_L$ as a function of $M$ is due to the
stability of the sliding mode.

To understand the nontrivial interplay of $F$, $M$, and $L_x$ in more
detail, we identified from the simulations three different paths,
(a), (b), and (c), dominating the flipping rate $\gamma$.  They
all describe the transition from a stable mode at the left wall
($\phi_s,x_s$) to another at the right wall ($\phi_s+\pi,L_x-x_s$) due
to fluctuations in the rod orientation $\phi$, whereas the
translational motion just follows the internal force $F$ and the
confining potential $V({\bf r},\phi)$~\cite{fluctuation_assumption}.
These three different paths are sketched in
Fig.~\ref{fig:turning_sketch} and are described as follows:
\begin{figure}
  \includegraphics[width=7cm]{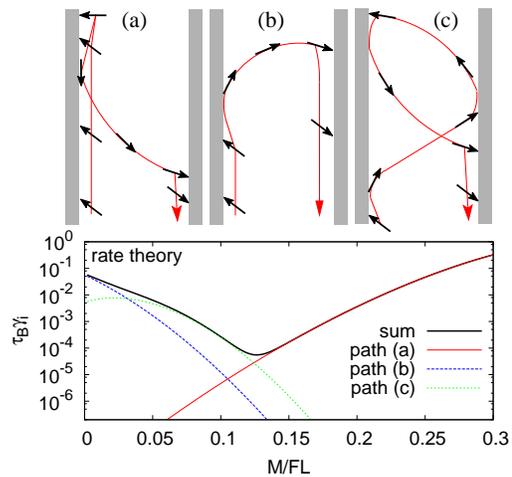}
  \caption{(Color online) Top panel: the paths governing the flipping rate: (a)
    turning in the direction of the torque, (b) turning against the
    direction of the torque, (c) three-stage event---first, turning
    in the direction of the torque and then turning against it. Bottom
    panel: The rates of the three different paths as a function of
    $M/LF$ for $\beta FL=60$, $L_x=8L$.}
  \label{fig:turning_sketch}
\end{figure}
The rod can slip out of its stable sliding mode by fluctuating in the
direction of the torque [path (a)] or by fluctuating against it [path
(b)]. In path (a), detachment from the (left) wall, which amounts to
overcoming a barrier in the torque/angle from $\phi_s$ to $\phi_u$, most
likely also leads to finding the stable mode on the other (right) wall (for
$L_x\lesssim R$). Path (b), however, is only successful if the rod
orientation is subject to strong and fast fluctuations which enable it to make
a turn of an angle $(-\pi+\phi_u-\phi_s)$ before reaching the other wall. This
explains why for intermediate torques and small $L_x$, another important
three-stage path (c) is dominating.  This path is initiated by a small
fluctuation of the orientation against the direction of the torque, from
$\phi_s$ toward $\pi/2$ on the (left) wall. In a second stage, the swimmer
approaches the other (right) wall at a small, constant turning velocity
$\dot{\phi}$, reaching it after only a short time due to its strong internal
force. By the other (right) wall it is reoriented in an upward direction
before, in a third stage, turning quickly in the direction of the torque such
that it reaches the original (left) wall at an angle $\phi_u$.  The flipping
rate is now given by the path integral $\gamma\propto\int D\phi\exp(-\beta
S[{\bf r},\phi]/4)$, keeping initial and final configurations of $\phi$ and
$x$ appropriately fixed.  Here, the Onsager-Machlup action is given by $S[{\bf
  r},\phi]=\int_0^\infty {\mathrm d}t^\prime\, \left|\partial_{t'}\phi(t') - M
+\partial_\phi V\left({\bf r}(t'),\phi(t')\right)\right|^2$~\cite{Onsager:53,
  footnote_swimmer_taniguchi}, with $t'=\beta D_r t$ the normalized time.
Note that our system is non-Hamiltonian due to the internal driving force and
the translation-rotation coupling. Hence, the least action path cannot be
found as the minimum energy path in some energy landscape, as vastly studied
in the literature~\cite{Vanden-Eijnden:08,Olender:97}. In contrast, we now
construct a non-Hamiltonian rate theory by assuming that---in the limit of
large forces $\beta FL$---the flipping rate $\gamma$ is dominated by either of
the three paths [$i=({\rm a}), ({\rm b}), ({\rm c})$], identified in the simulation.  The
respective minimum actions are given by $S_i[{\bf r}_i,\phi_i]$, with
$\phi_i(t')$ minimizing the action subject to the constraints
[$\phi_i(0)=\phi_s, x_i(0)=x_s$] and [$\phi_i(\infty)=\phi_s\pm\pi,
x_i(\infty)=L_x-x_s$], where the plus sign corresponds to paths (a) and
(c), and the minus sign to path (b).  Paths (a) and (b) [(c)]
are further constrained by the condition not to reach (to reach) the
initial wall between the initial and the final stage.

In order to calculate the associated actions for the different paths,
we divide the trajectories into parts where the front rod tip sits on
the original (left) wall and into parts where the rod moves at a
constant turning velocity $\dot{\phi}_i$ in between the walls. The
former parts can then be expressed as the barrier heights
$4\int_{\phi_s}^{\phi_m} {\mathrm d}\phi |M-\partial_\phi
V|$~\cite{Olender:97}, with $\phi_m=\phi_u$ for path (a) and
$\phi_m=\pi/2$ for paths (b), (c), whereas the latter are simply
given by $|\partial_{t'}\phi_i-M|^2t^\prime_{\max}$, $t^\prime_{\max}$
being the normalized time it takes to swim from one wall to the other
($t^\prime_{\max}$ is chosen to minimize the action). The individual
rates are roughly given by $\gamma_i\approx\exp[-\beta S_i/4]/\tau_B$,
where the kinetic prefactors are crudely approximated by $1/\tau_B$,
and plotted in Fig.~\ref{fig:turning_sketch}.  Summation over the
individual rates yields the long-time diffusion coefficient
$D_L\approx v_y^2\gamma^{-1}$, with $\gamma\simeq \sum_i\gamma_i$
plotted as a function of $M$ for different $L_x$ in
Fig.~\ref{fig:longtimediffusion_confinement_bulk}(b). The rate theory
reproduces clearly the $L_x$ dependence and the nonmonotonicity of
$D_L$ as a function of $M$ and attributes it to different rates of the
paths (a) and (c). Moreover, the maximum in $D_L$ is predicted to
be weakly dependent on $L_x$ in agreement with the simulations.
However, the actual values of the rate theory differ from the
simulation data due to the crude approximation made for the kinetic
prefactors.

Finally, we study the effect of an additional gravitational field in
the $x$ direction ($k>0$), breaking the symmetry of the channel
potential (see the right inset of
Fig.~\ref{fig:msd_confinement_gravity}).  On average, the swimmer is
now situated more on the left than on the right channel wall, such
that the sliding mode becomes {\it ballistic} (see
Fig.~\ref{fig:msd_confinement_gravity}).

In conclusion, we have studied the dynamic behavior of a
self-propelled Brownian rod performing circular motion.  In the bulk,
the analytical solution reveals long-time diffusive behavior. In
channel confinement, an efficient stable sliding mode was identified
that strongly enhances the long-time diffusion along the channel as
obtained by computer simulation and a non-Hamiltonian rate theory.  If
the channel is asymmetric, the sliding mode leads to ballistic
long-time motion.

The sliding motion of circle swimmers can be verified in experiments
with different set-ups: First, catalytically driven
nanorods~\cite{Dhar:06,Walther:08} and self-propelled magnetic
colloidal rods confined to a microchannel~\cite{Koppl:06} will exhibit
sliding~\cite{Merkt:06}.  Second, confined
bacteria~\cite{Berg:90,Hill:07,DiLuzio:05} and
spermatozoa~\cite{Riedel:05,Woolley:03} move in two dimensions along
circles.  In fact, the typical radius of the observed circular motion
is in the range of $10-1 000\mu {\rm m}$ for
spermatozoae~\cite{Riedel:05,Woolley:03} and of the order of $50\mu {\rm m}$
for Escherichia coli bacteria~\cite{Berg:90}.  Therefore, the radii are
typically larger but comparable with the particle sizes. When these
particles are exposed to microchannels of similar widths as the
observed radii, as realized for the bacteria~\cite{DiLuzio:05}, the
predicted huge acceleration behavior should be observed, as has
already been seen in 3D~\cite{Berg:90}.  Third, vibrated polar
granular rods~\cite{Kudrolli:07} with an additional left-right
asymmetry perform circle motions. When placed into a slit geometry, a
sliding effect may be observed here as well.

Accelerating the dynamics in the channel by tuning the torque may be
exploited as a mechanism to separate a certain species out of a
crowded solution of different active particles. If a microfluidic
channel is connected to a bulk mixture, the species moving quickest
along the channel will arrive first at the channel end and can
efficiently be removed. This might be more efficient than traditional
separation techniques such as capillary
electrophoresis~\cite{Rodriguez:04}.

%\begin{acknowledgments} 
  We thank U.\ Zimmermann, H.\ H.\ Wensink, A.\ Wynveen and W.\ C.\
  K.\ Poon for helpful discussions.  This work has been supported by
  the DFG through the SFB TR6 (project C3).
%\end{acknowledgments}



\begin{thebibliography}{26}
\bibitem[{flu()}]{fluctuation_assumption}
\bibinfo{note}{This assumption is clearly justified for large ratios
  $D_r/D_\parallel$, $D_r/D_\perp\gg 1/L^2$, and $FL/M\gg 1$.}
\end{thebibliography}
\end{document}